\documentclass[preprint,12pt]{elsarticle}

\usepackage{amssymb}
\usepackage{amsthm}
\usepackage[fleqn]{amsmath}
\usepackage{graphicx}
\usepackage{caption}
\usepackage{subcaption}
\usepackage{float}
\usepackage{mathtools}
\usepackage{mathrsfs}
\usepackage{lineno}
\usepackage{setspace}
\begin{document}

\title{Can We Remove Secular Terms for Analytical Solution of Groundwater Response under Tidal Influence?}
\author{Selva Balaji Munusamy\fnref{Munusamy S.B.}}
\ead{selva191@gmail.com}
\fntext[Munusamy S.B.]{Research Scholar, Department of Civil Engineering, Indian Institute of Technology Kharagpur, Kharagpur WB 721302, India}
\author{Anirban Dhar\corref{cor1}\fnref{Dhar A.}}
\ead{anirban.dhar@gmail.com, anirband@civil.iitkgp.ernet.in}
 \fntext[Dhar A.]{Assistant Professor, Department of Civil Engineering, Indian Institute of Technology Kharagpur, Kharagpur WB 721302, India}
 \cortext[cor1]{Tel.: +91 3222 283432 (O), +91 3222 283433 (R); fax: +91 3222 282254}

 \begin{frontmatter}



%%\title{}


%%\author{}

%%\address{}

\begin{abstract}
\noindent This paper presents a secular term removal methodology based on the homotopy perturbation method for analytical solutions of nonlinear problems with periodic boundary condition. The analytical solution for groundwater response to tidal fluctuation in a coastal unconfined aquifer system with the vertical beach is provided as an example. The non-linear one-dimensional Boussinesq's equation is considered as the governing equation for the groundwater flow. An analytical solution is provided for non-dimensional Boussinesq's equation with cosine harmonic boundary condition representing tidal boundary condition. The analytical solution is obtained by using homotopy perturbation method with a virtual embedding parameter.  The present approach does not require pre-specified perturbation parameter and also facilitates secular terms elimination in the perturbation solution. The solutions starting from zeroth-order up to third-order are obtained. The non-dimensional expression, $A/D_{\infty}$ emerges as an implicit parameter from the homotopy perturbation solution. The non-dimensional solution is valid for all ranges of $A/D$ values. Higher order solution reveals the characteristics of the tidal groundwater table fluctuations.

\end{abstract}


\end{frontmatter}

%\linenumbers
\section*{Notation}
\nopagebreak
\par
\begin{tabular}{r  @{\hspace{1em}=\hspace{1em}}  l}
$()_{,x}$ ,$()_{,xx}$, $()_{,t}$             & $\partial^2 ()/\partial x^2$, $\partial ()/\partial x$, $\partial ()/\partial t$;\\
$h$                    & local water table height (m);\\
$x$                    & longitudinal coordinate (m);\\
$X$                    & non-dimensional longitudinal coordinate (m);\\
$t$                    & time (s);\\
$T$                    & non-dimensional time (s);\\
$p$                    & embedding virtual parameters;\\
$A$                    & amplitude of the tidal oscillations (m);\\
$D$                    & mean sea level above aquifer bottom (m); \\
$D_{\infty}$           & thickness of unconfined aquifer (m);\\
$H$                    & $h/D_{\infty}$;\\
$H_0$, $H_1$, $\dots$  & Perturbation expansion terms of $H$;\\
$J_{1}$, $J_{2}$       & inhomogeneous terms;\\
$K$                    & hydraulic conductivity of the aquifer (m/s);\\
$L_{\infty}$           & $\sqrt{D_{\infty} K/(\eta_e \omega)}$ (m);\\
$\eta_e$               & effective porosity of the aquifer;     \\
$\omega$               & angular frequency of the tide (1/s).
\end{tabular}

\section{Introduction}

\noindent Groundwater head fluctuations due to tidal influences are an important phenomenon to understand seawater intrusion process. The interaction between coastal tide and groundwater table is non-linear due to non-linear filtering effects of beach sand \cite{Nielsen1990}. The tidal water fluctuation propagates inside coastal aquifer with attenuated amplitude and phase lag due to the nonlinearity of the aquifer material. The interaction between tides and groundwater can be reasonably modeled by using non-linear Boussinesq's equation with Dupuit's assumption.  \cite{Dagan1967} derived Boussinesq's equation for shallow water approximation and provided a perturbation solution for vertical beach face. \cite{Parlange1984} derived first-order perturbation (starting from zeroth-order) solutions on the basis of \cite{Dagan1967} by using a shallow water perturbation parameter. Parlange's solution showed an increase in oscillation due to the second sub-harmonics in the first-order solution (with double the speed of tidal frequency) and water table height. However, their solution stopped at first-order for their approach could not handle secular terms in higher orders. \cite{Nielsen1990} provided a generalised perturbation solution for small amplitude based on linearized Boussinesq's equation for the vertical beach as well as for a sloped beach face without considering density-dependent flow. \citet{Sun1997} derived a two-dimensional solution for transient groundwater flow in a confined aquifer in an estuary by using two-dimensional groundwater flow equation with damped cosine function as a seaside boundary condition. \citet{Li2000} provided a solution for spring-neap tidal water table fluctuations in a sloped coastal aquifer considering two different tidal constituents by using linearized Bousssinesq's equations with same parameters as \citet{Nielsen1990}. \citet{Teo2003} provided a solution for groundwater fluctuations in a coastal aquifer with sloping beach based on amplitude parameter ($\alpha=A/D$) and shallow water parameter ($\epsilon=\sqrt{\eta_e \omega D/(2 K)}$). However, the solution is restricted to small values of shallow water parameter ($\epsilon\ll1$).  \citet{Jeng2005} perturbation solution for two-dimensional tide-induced water table fluctuation for sinusoidal coastal line without considering moving boundary effects as the origin fixed at mean tide level. \citet{Song2007} and \citet{Kong2011} used maximum time-averaged water table height and thickness of the unconfined aquifer, which included tide-induced overheight respectively to get perturbation parameter which is always lesser than one to get a higher order solution for vertical beach face and sloped beach respectively. In perturbation solutions of non-linear equations, secular terms generate unbounded solutions.  \cite{Song2007} and \cite{Kong2011} have showed that the secular terms can be removed from the boundary conditions by using series expansions. However, pre-specification of small perturbation parameters is essential for classical perturbation approach. \cite{Stojsavljevic2012} provided a semi-analytical solution for groundwater fluctuation due to tidal boundary condition based on the perturbation parameters introduced by \cite{Teo2003}. \cite{Stojsavljevic2012} proposed a solution form based on coefficients. The solution is not valid for higher values of $\epsilon$. The analytical solutions can be used for the inverse modelling problem to fit the parameters for wave propagation in coastal aquifers \cite{Raubenheimer1999}.

\noindent Homotopy perturbation method overcomes the problem of selecting parameter(s) for the solution by embedding a virtual parameter ($p$) in the homotopy equation which is virtually small ($p\leq1$) . In the present study homotopy perturbation method has been utilized to get a higher order analytical solution of water table fluctuation driven by tidal forcing for an unconfined aquifer with a vertical beach face. This method does not require any pre-specified perturbation parameters. Secular terms have been removed from the solution by using series expansions for the boundary conditions, similar to \cite{Song2007} and \cite{Kong2011}. The solution is not restricted for any parameter values.

\section{Methodology}

\subsection{Problem setup}

\noindent One-dimensional tidal groundwater flow is considered for an unconfined aquifer with a long, straight and vertical beach face. The aquifer bottom considered to be a straight horizontal impervious bottom. The aquifer hydraulic conductivity($K$) is assumed to be homogeneous, isotropic. Regional freshwater flow towards the sea, vertical flows (Dupuit\textemdash Forchheimer assumption) and capillary effects are assumed to be negligible. There is no decoupling between the seawater level and the water table at the seaside boundary. Groundwater flow equation in a long straight beach with homogeneous and isotropic hydraulic conductivity can be written as (Boussenisq's equation),
\begin{align}\label{eqn:GE}
h_{,t}&={K}/{\eta_e}\:(h\:h_{,x})_{,x}
\end{align}
\noindent Where, $h(x,t)$ is local water table height, $t$ is time, $x$ is the distance from the intersection of mean sea level and beach, $\eta_e$ is the effective porosity of the aquifer, $D$ is mean sea level from aquifer bottom, $h_{,t}$ is partial differential of h with respect to $t$ ($\partial h/\partial t$), and $h_{,x}$ is partial differential of h with respect to $x$ ($\partial h/\partial t$). In Boussenisq's equations (\ref{eqn:GE}) right-hand side terms introduce non-linearity in the problem. The height of the water table at the beach face boundary considering only one tidal constituent with tidal amplitude $A$, and angular frequency $\omega$ can be represented in the following form,
\begin{align}\label{eqn:BC}
h(0,t)&= D+A \cos[\omega t]
\end{align}
\noindent The initial condition of water table is given as,
\begin{align}\label{eqn:IC}
h(x,0)&= D
\end{align}
\noindent The inland boundary condition is consistent with the assumption that water table fluctuations at an asymptotic distance ($x \rightarrow \infty$) from seaside boundary would be negligible.
\begin{align} \label{eqn:BCR}
\left.h_{,x}\right\vert_{x \to \infty}&=0
\end{align}

\subsection{Homotopy perturbation Approach}

\noindent \cite{He1999} proposed homotopy perturbation method by coupling perturbation method with homotopy technique. Homotopy perturbation method overcomes limitations exist due to the assumption of the small perturbation parameter in the problem and, selection and determination of parameter. General non-linear differential equation can be given as,
\begin{align}
L(u)+N(u)=0
\end{align}
\noindent Where, L and N are linear and nonlinear operators respectively. According to the standard homotopy method, we can construct a homotopy equation of the following form,
\begin{align}
\mathscr{H}(u,p)=\tilde{L}(u)+p\left[N(u)+L(u)-\tilde{L}(u)\right]=0\label{eqn:SHPM_construct}
\end{align}
\noindent Where, $p$ is a small virtual parameter in the homotopy equation for perturbation solution, $\tilde{L}$ is a linear operator and $\tilde{L}(u)=0$ can approximately describe the solution property. The virtual parameter $p$ monotonically increased from zero to unity as the approximate problem [$\mathscr{H}(u,0)$] is continuously deformed to original one [$\mathscr{H}(u,1)$]. In topology, this is called as homotopic between two functions, and the functions $\tilde{L}(u)$ and $N(u)+L(u)-\tilde{L}(u)$ are called as homotopic. By using the perturbation parameter $p$, the solution can be assumed in the perturbation form similar to traditional perturbation methods as follows,
\begin{align}
u=\sum_{n=0}^{\infty}p^n\:u_n \label{eqn:uinfty_series}
\end{align}
\noindent The perturbation solution can be obtained by replacing $p=1$ in the equation (\ref{eqn:uinfty_series}), higher-order solution can be obtained in the following form,
\begin{align}
u=\lim_{p \to 1}\sum_{n=0}^{\infty}p^n\:u_n=u_0+u_1+u_2+\dots
\end{align}
\noindent By substituting the above perturbation solution into the equation (\ref{eqn:SHPM_construct}), the standard homotopy perturbation equation can be divided into linear parts by comparing the coefficients of $p^n$ for $n=0,1,\dots$, $n$. The solutions for $u_0$, $u_1$, $u_2$, $\dots$ are obtained by solving the decomposed linear differential equations. The value of $n$ is dependent on the required accuracy.

\section{Application of the Homotopy Perturbation Method}

\subsection{Analytical Solution of the Non-dimensional Equation}

\noindent The Boussinesq's equation can be non-dimensionalized by using the following mappings: $X\rightarrow x/L_{\infty}$, $H\rightarrow h/D_{\infty}$, $T\rightarrow\omega t$, $L_{\infty}\rightarrow \sqrt{D_{\infty} K/(\eta_e \omega)}$. \cite{Kong2011} have also used $D_{\infty}$ for non-dimensionalizing the height of water table. The non-dimensional Boussinesq's equation can be given as,

\begin{align}
H_{,T}&=(H\:H_{,X})_{,X}\label{eqn:GE_ND}
\end{align}
\noindent Boundary conditions for the non-dimensional problem can be given as,
\begin{align}
H(0,T)&=D/D_{\infty}+[A\:\cos(T)]/D_{\infty}\label{eqn:BC_ND}\\
\left.H_{,X}\right\vert_{X \to \infty}&=0
\end{align}
\noindent The homotopy equation for non-dimensional groundwater flow equation can be written by using small virtual embedding parameter $p$ as follows,
\begin{align}
H_{,T}-H_{,XX}+p[-(H\:H_{,X})_{,X}+H_{,XX}]&=0\label{eqn:HPM_ND}
\end{align}
\noindent Where, $H_{,T}$ is $\partial H/\partial T$,$H_{,X}$ is $\partial H/\partial X$, $L(h)=H_{,T}$, $\tilde{L}(H)=H_{,T}-H_{,XX}$, $N(H)=-(H\:H_{,X})_{,X}$. The second term in the equation (\ref{eqn:HPM_ND}), $H_{,XX}$ ($=\partial^2 H/\partial X^2$) approximates the nonlinear diffusion term in the Boussinesq's equation for small amplitude oscillations compared to $D$. To stop the generation of secular terms in the particular integral, boundary condition can be written in series form as follows \citep{Song2007,Kong2011},
\begin{align}
H(0,T)&=[A\:\cos(T)]/D_{\infty}+\underbrace{1+\sum_{n=1}^{\infty} b_n p^n}_{D/D_{\infty}}\label{eqn:BCND_perturabtion}
\end{align}
\noindent By substituting the equation (\ref{eqn:uinfty_series}) into the equations (\ref{eqn:HPM_ND}) and (\ref{eqn:BCND_perturabtion}) with $u\equiv H$, the perturbation equations along with boundary conditions for the different powers of virtual parameter can be written as,
\begin{equation}
\begin{aligned}\label{eqn:p0ND}
p^0:
\begin{dcases}
H_{0,T}-H_{0,XX}=0\\
H_0(X,T) = [A\:\cos(T)]/D_{\infty}+1\\
\left.H_{0,X}\right\vert_{X \to \infty}=0
\end{dcases}
\end{aligned}
\end{equation}
\begin{equation}
\begin{aligned}\label{eqn:p1ND}
p^1:
\begin{dcases}
H_{1,T}- H_{1,XX}=\underbrace{\left[ (H_0\: H_{0,X})_{,X}-\:H_{0,XX}\right]}_{J_1}\\
H_1(X,T) = b_1\\
\left.H_{1,T}\right\vert_{X \to \infty}=0.
\end{dcases}
\end{aligned}
\end{equation}
\begin{equation}
\begin{aligned}\label{eqn:p2ND}
p^2:
\begin{dcases}
H_{2,T}- H_{2,XX}=\underbrace{\left[ (H_1\: H_{0,X}+H_0\: H_{1,X})_{,X}-\:H_{1,XX}\right]}_{J_2}\\
H_2(X,T) = b_2\\
\left.H_{2,X}\right\vert_{X \to \infty}=0.
\end{dcases}
\end{aligned}
\end{equation}
\begin{equation}
\begin{aligned}\label{eqn:p3ND}
p^3:
\begin{dcases}
H_{3,T}- H_{3,XX}\\
=\underbrace{\left[ (H_2\: H_{0,X}+H_1\: H_{1,X}+H_0\: H_{2,X})_{,X}-\:H_{2,XX}\right]}_{J_3}\\
H_3(X,T) = b_3\\
\left.H_{3,X}\right\vert_{X \to \infty}=0.
\end{dcases}
\end{aligned}
\end{equation}
\noindent where, $J_1$, $J_2$, and $J_3$ are inhomogeneous terms. The parameters $b_1$, $b_2$, $b_3$, $\dots$ can be determined from the resulting solutions. Solution for zeroth-order equation (\ref{eqn:p0ND}) is given by,
\begin{align}
H_{0}(X,T)&=1+\frac{A}{D_{\infty}}\:e^{-X/\sqrt{2}}\:\cos[T-X/\sqrt{2}]\label{eqn:0th_NDsolution}
\end{align}
\noindent By substituting $H_{0}(X,T)$ (\ref{eqn:0th_NDsolution}) in the equation (\ref{eqn:p1ND}), inhomogeneous terms in the first-order equation can be obtained as,
\begin{align}
J_1&=\frac{A^2}{D_{\infty}^2}\left\{\frac{1}{2}\:e^{-\sqrt{2} X}-e^{-\sqrt{2} X}\sin[2 T -\sqrt{2} X]\right\}\label{eqn:J1}
\end{align}
\noindent Solution for the first-order equation ($H_1$) is given by,
\begin{align}
H_{1}(X,T)&=\frac{A^2}{D_{\infty}^2}\left\{-\frac{1}{4}\:e^{-\sqrt{2}X}+\frac{1}{2}\:e^{-X}\cos[2 T-X]\right.\nonumber\\
&\quad\left.-\frac{1}{2}\:e^{-\sqrt{2}X}\:\cos[2 T-\sqrt{2}X]\right\}\label{eqn:1st_NDsolution}
\end{align}
\noindent The term ${A^2}/({4 D_{\infty}^2})$ in the complementary function of first order solution ($H_{1}^{CF}$) generates secular term in the second order inhomogeneous term ($J_2$). By applying left boundary condition for the requirement of no secular term in second-order inhomogeneous term, $b_1$ can be given as,
\begin{align}
b_1=-\frac{A^2}{4 D_{\infty}^2}
\end{align}
\noindent The particular integral $H_{P.I}(X,T)$ becomes unbounded ${\rm numerator}/{0}$ for the secular terms (see Appendix),
\begin{align}
H^{PI}(X,T)=\frac{1}{f(\Delta,\Delta')}g(X,T)=\frac{Numerator}{0}
\end{align}
\noindent By substituting (\ref{eqn:1st_NDsolution}) and (\ref{eqn:0th_NDsolution}) into the equation (\ref{eqn:p2ND}), the inhomogeneous terms obtained as,
\begin{align}
J_2&=\frac{A^3}{D_{\infty}^3}\left\{-2\:e^{-\frac{3}{\sqrt{2}} X}\:\cos[T-\frac{X}{\sqrt{2}}]\right.\nonumber\\
&\quad+\frac{1}{\sqrt{2}}\:e^{-\frac{1}{2}(2+\sqrt{2})X}\:\cos[T+\frac{1}{2}(-2+\sqrt{2})X]\nonumber\\
&\quad+\frac{9}{4}\:e^{-\frac{3}{\sqrt{2}}X}\:\sin[3 T-\frac{3}{\sqrt{2}}X]\nonumber\\
&\quad+\frac{3}{2}\:e^{-\frac{3}{\sqrt{2}}X}\:\sin[T-\frac{1}{\sqrt{2}}X]\nonumber\\
&\quad-\frac{1}{4}(3+2\sqrt{2})\:e^{-\frac{1}{2}(2+\sqrt{2})X}\:\sin[3 T-\frac{1}{2}(2+\sqrt{2})X]\nonumber\\
&\quad\left.-\frac{1}{4}\:e^{-\frac{1}{2}(2+\sqrt{2})X}\:\sin[ T+\frac{1}{2}(-2+\sqrt{2})X]\right\}\nonumber\\
&\quad-\frac{A^3}{4 D_{\infty}^3}\:\underbrace{e^{-\frac{1}{\sqrt{2}}X}\:\sin[T-\frac{1}{\sqrt{2}}X]}_{secular\:term}\label{eqn:J2}
\end{align}
\begin{align}
b_2&=0
\end{align}
\noindent The secular term is removed by defining $b_1=-{A^2}/({4 D_{\infty}^2})$ in the first-order solution. Solution for second-order equation ($H_2$) is given by,
\begin{align}
&H_{2}(X,T)\nonumber\\
&=\frac{A^3}{D_{\infty}^3}\left\{\frac{1}{16}(-2+3\sqrt{2})\:e^{-\sqrt{\frac{3}{2}} X}\:\cos[3 T-\sqrt{\frac{3}{2}}X] \right.\nonumber\\
&\quad+\frac{3}{8}\:e^{-\frac{3}{\sqrt{2}}X}\:\cos[3 T-\frac{3}{\sqrt{2}}X]\nonumber\\
&\quad+\frac{11}{20}\:e^{-\frac{3}{\sqrt{2}}X}\:\cos[ T-\frac{1}{\sqrt{2}}X]\nonumber\\
&\quad-\frac{3}{10}\:e^{-\frac{1}{\sqrt{2}}X}\:\cos[T-\frac{1}{\sqrt{2}}X]\nonumber\\
&\quad-\frac{1}{4}\:e^{-\frac{1}{2}(2+\sqrt{2})X}\:\cos[T+\frac{1}{2}(-2+\sqrt{2})X]\nonumber\\
&\quad-\frac{1}{16}(4+3\sqrt{2})\:e^{-\frac{1}{2}(2+\sqrt{2})X}\:\cos[3 T-\frac{1}{2}(2+\sqrt{2})X]\nonumber\\
&\quad-\frac{1}{10}\:e^{-\frac{3}{\sqrt{2}}X}\:\sin[T-\frac{1}{\sqrt{2}}X]\nonumber\\
&\quad-\frac{1}{80}(-8+5\sqrt{2})\:e^{-\frac{1}{\sqrt{2}}X}\:\sin[T-\frac{1}{\sqrt{2}}X]\nonumber\\
&\quad\left.+\frac{1}{8\sqrt{2}}\:e^{-\frac{1}{2}(2+\sqrt{2})X}\:\sin[T+\frac{1}{2}(-2+\sqrt{2})X]\right\}
\end{align}
\noindent The parameter $D_{\infty}$ can be determined by solving the implicit equation,
\begin{align}
\frac{D}{D_{\infty}}&=1+b_1+b_2+b_3\label{eqn:sol_dinfty}
\end{align}

\noindent The inhomogeneous terms in the third-order equation, third-order solution, and $b_3$ are provided in the Appendix. From the solution, it can be observed that the physical parameters involved in the governing equation generate an implicit parameter $A/D_{\infty}$ for the solution.

\subsection{Convergence}

\begin{figure}
\centering
    \includegraphics[width=\linewidth]{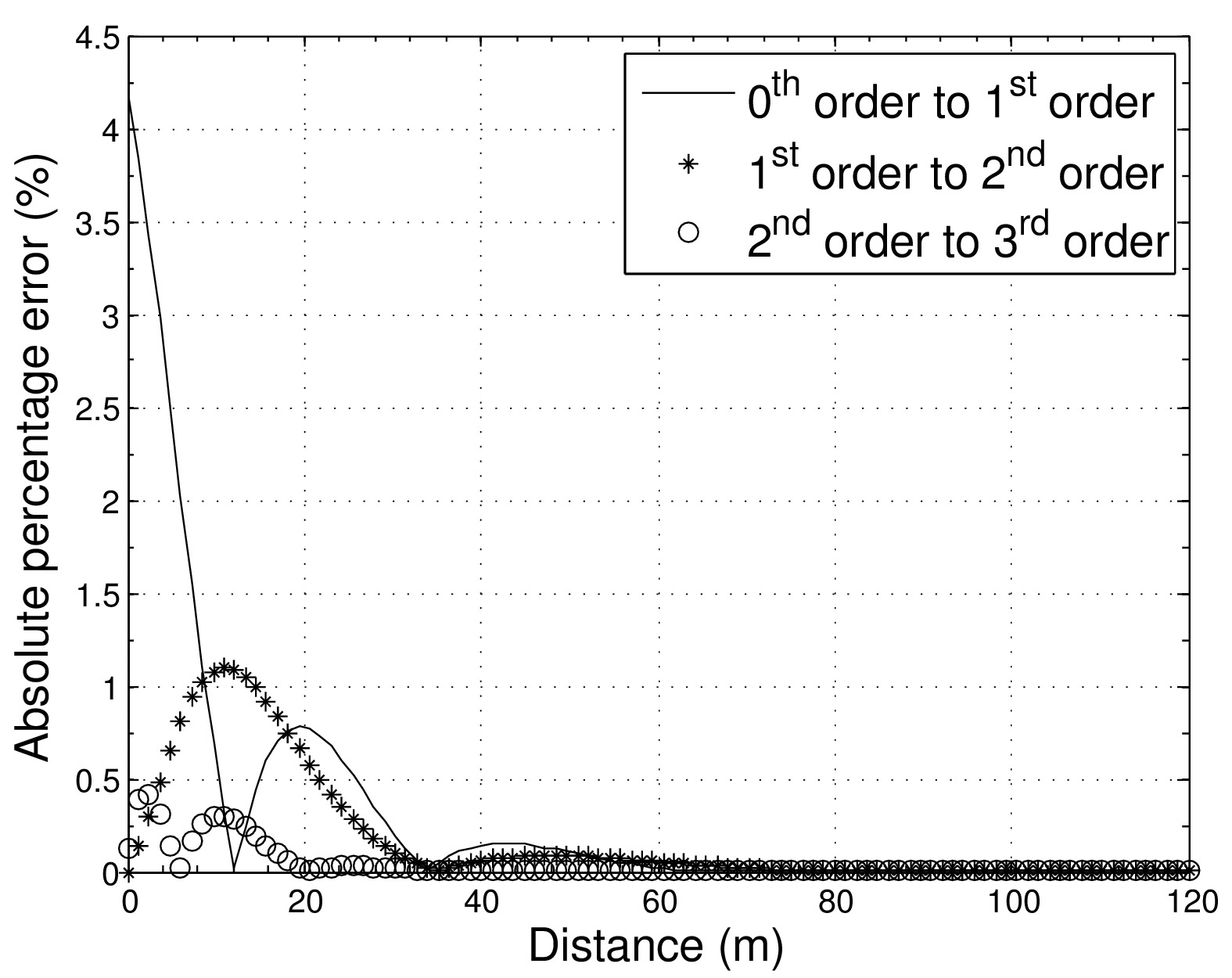}
    \caption{ \small Convergence of homotopy perturbation solution up to third-order by using maximum absolute percentage change for an coastal unconfined aquifer with field scale parameters: a tidal wave with timeperiod= 12 hrs; D=5 m; A=4 m, $\eta_e=0.3$, $K=0.00089\:m/s$}
    \label{fig:per_change}
\end{figure}

\noindent The convergence of the proposed homotopy solution is analyzed by the plot of distance vs absolute maximum percentage error as shown in Figure \ref{fig:per_change}. The absolute maximum percentage error for different distances from the seaside boundary are obtained by calculating  maximum percentage change over the full tidal cycle at each point. The absolute percentage error can be calculated by using the following expression,
\begin{align}
&\varepsilon_{n_t,n_t-1}(x)\nonumber\\
&\quad=\max_{t} \left(\frac{\sum_{i=0}^{n_t} H_i(x,t)-\sum_{i=0}^{n_t-1} H_{i}(x,t)}{\sum_{i=0}^{n_t-1} H_{i}(x,t)}\times 100\right)
\end{align}
\noindent where, $\varepsilon_{n_t,n_t-1}(x)$ is the absolute maximum percentage error between the solution up to $n_t^{th}$ order and the solution up to $(n_t-1)^{th}$ over full tidal cycle at the distance x from the seaside boundary, $\sum_{i=0}^{n_t} H_i(x,t)$ is the solution up to $n_t^{th}$ order, and $\sum_{i=0}^{n_t-1} H_{i}(x,t)$ is the solution up to $(n_t-1)^{th}$ order. The maximum absolute percentage error between first-order and zeroth-order ($\varepsilon_{1,0}(x)$) is relatively very high while compared to the same of higher order solutions comparisons. The $\varepsilon_{1,0}(x)$ is highest at the seaside boundary with value of $4.2\%$ and decreases to zero at 12 m inland before increasing again to $0.75\%$ at 20 m inland. The maximum absolute percentage error between second-order and first-order $\varepsilon_{2,1}(x)$ is zero at the boundary and increases to its peak value $1.1\%$ at around 12 m inland. However, $\varepsilon_{2,1}(x)$ is small compared to $\varepsilon_{1,0}(x)$. The maximum absolute percentage error between third-order and second-order $\varepsilon_{3,2}(x)$ is very small with maximum value $0.48\%$ compared to lower order absolute errors. $\varepsilon_{3,2}(x)$ is zero from 20 m inland distance.

\subsection{Solution Comparisons and Discussions}

\begin{figure*}
        \centering
        \begin{subfigure}[b]{0.475\textwidth}
            \centering
            \includegraphics[width=\textwidth]{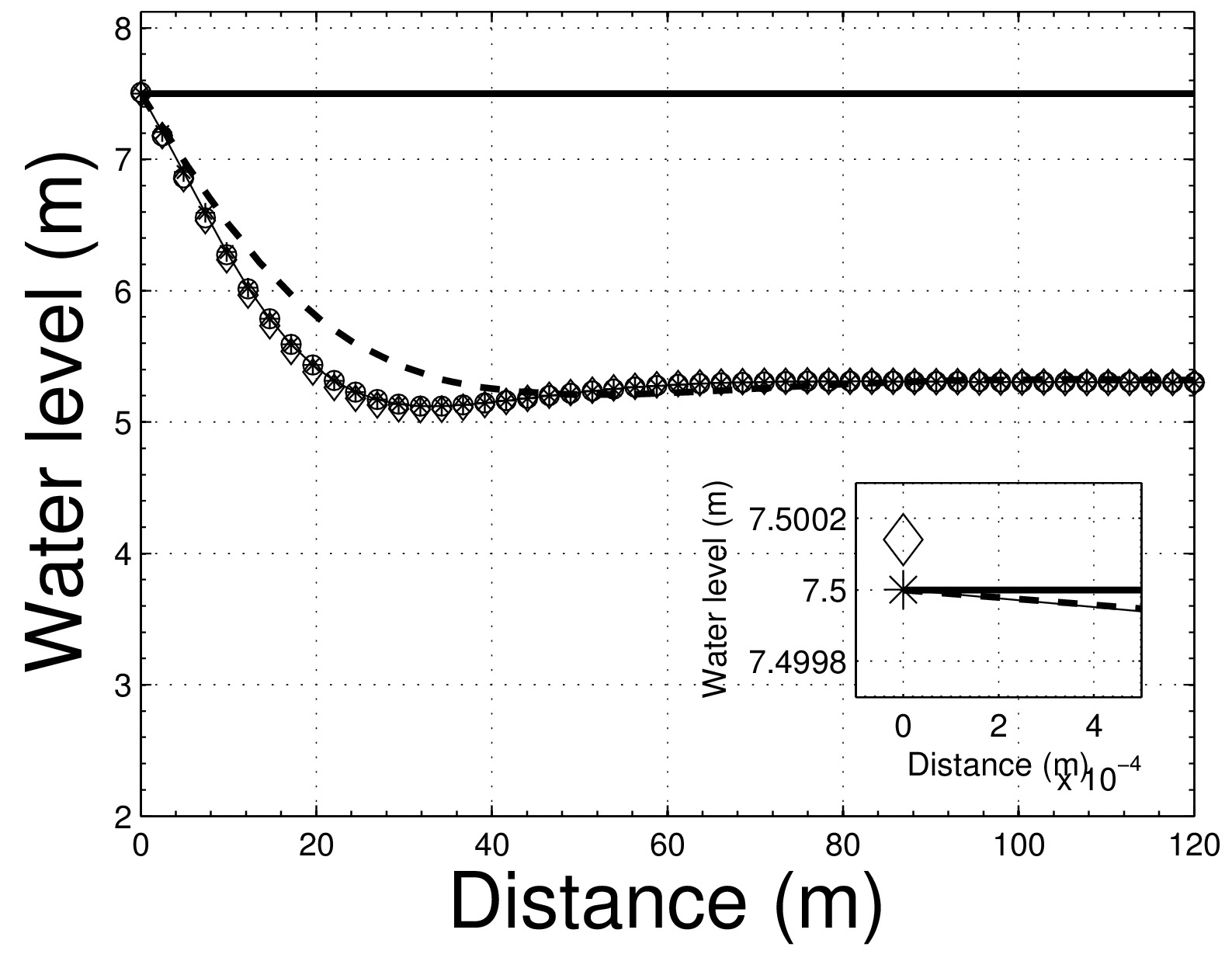}
            \caption[]%
            {{\small t=0 s }}
            \label{fig:t_0}
        \end{subfigure}
        \hfill
        \begin{subfigure}[b]{0.475\textwidth}
            \centering
            \includegraphics[width=\textwidth]{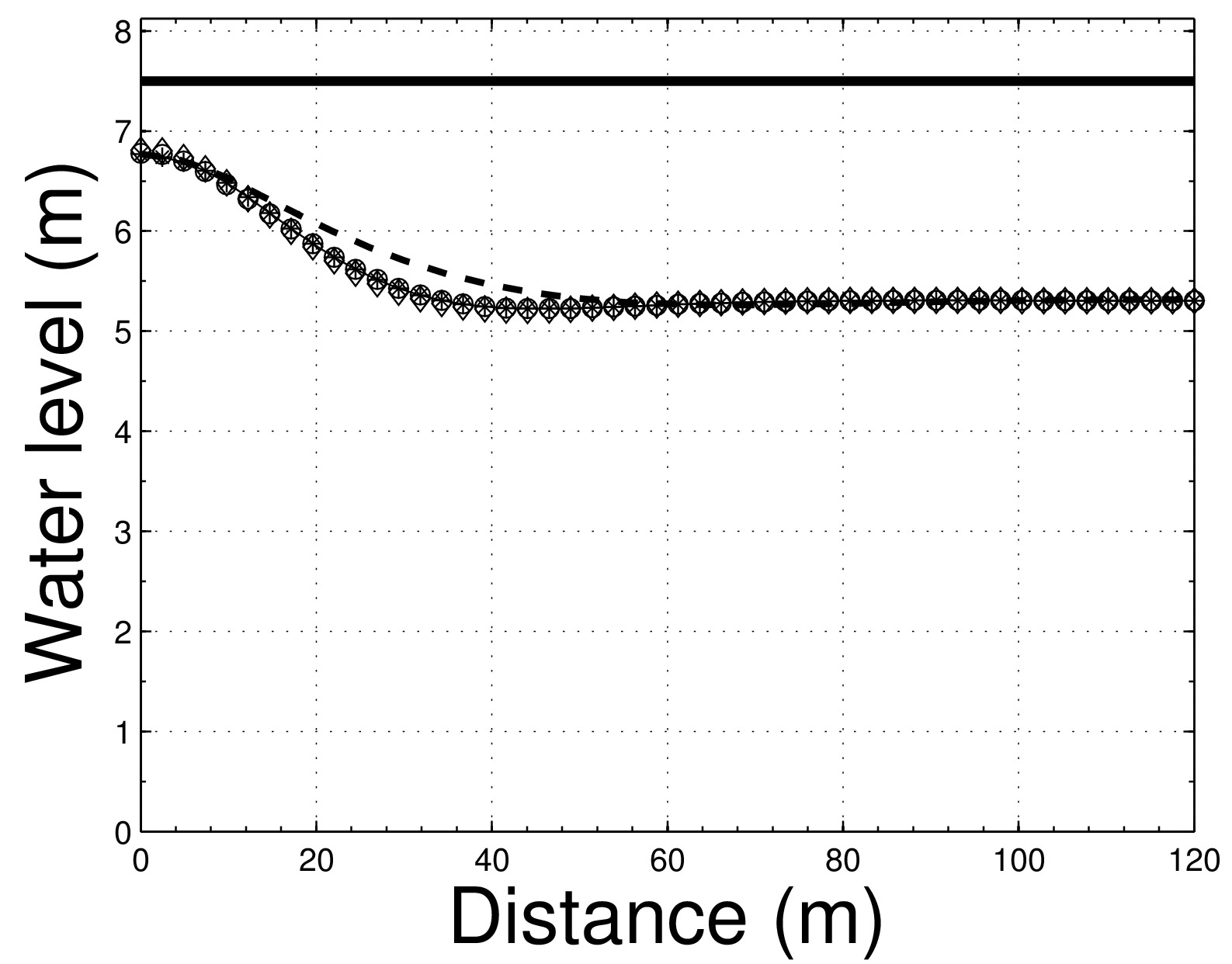}
            \caption[]%
            {{\small t=1.5 hrs  }}
            \label{fig:t_8}
        \end{subfigure}
        \vskip\baselineskip
        \begin{subfigure}[b]{0.475\textwidth}
            \centering
            \includegraphics[width=\textwidth]{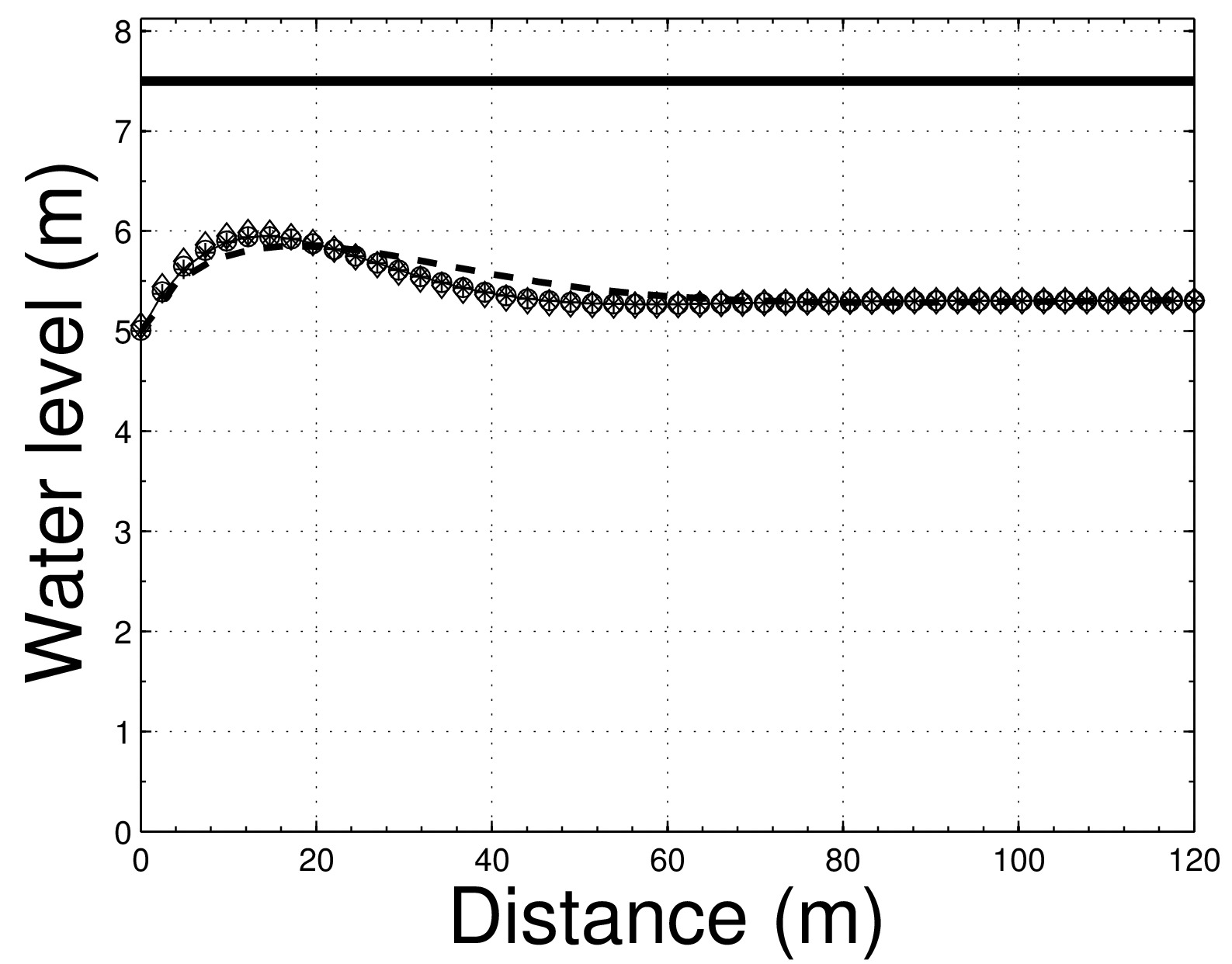}
            \caption[]%
            {{\small t=3 hrs}}
            \label{fig:t_16}
        \end{subfigure}
        \quad
        \begin{subfigure}[b]{0.475\textwidth}
            \centering
            \includegraphics[width=\textwidth]{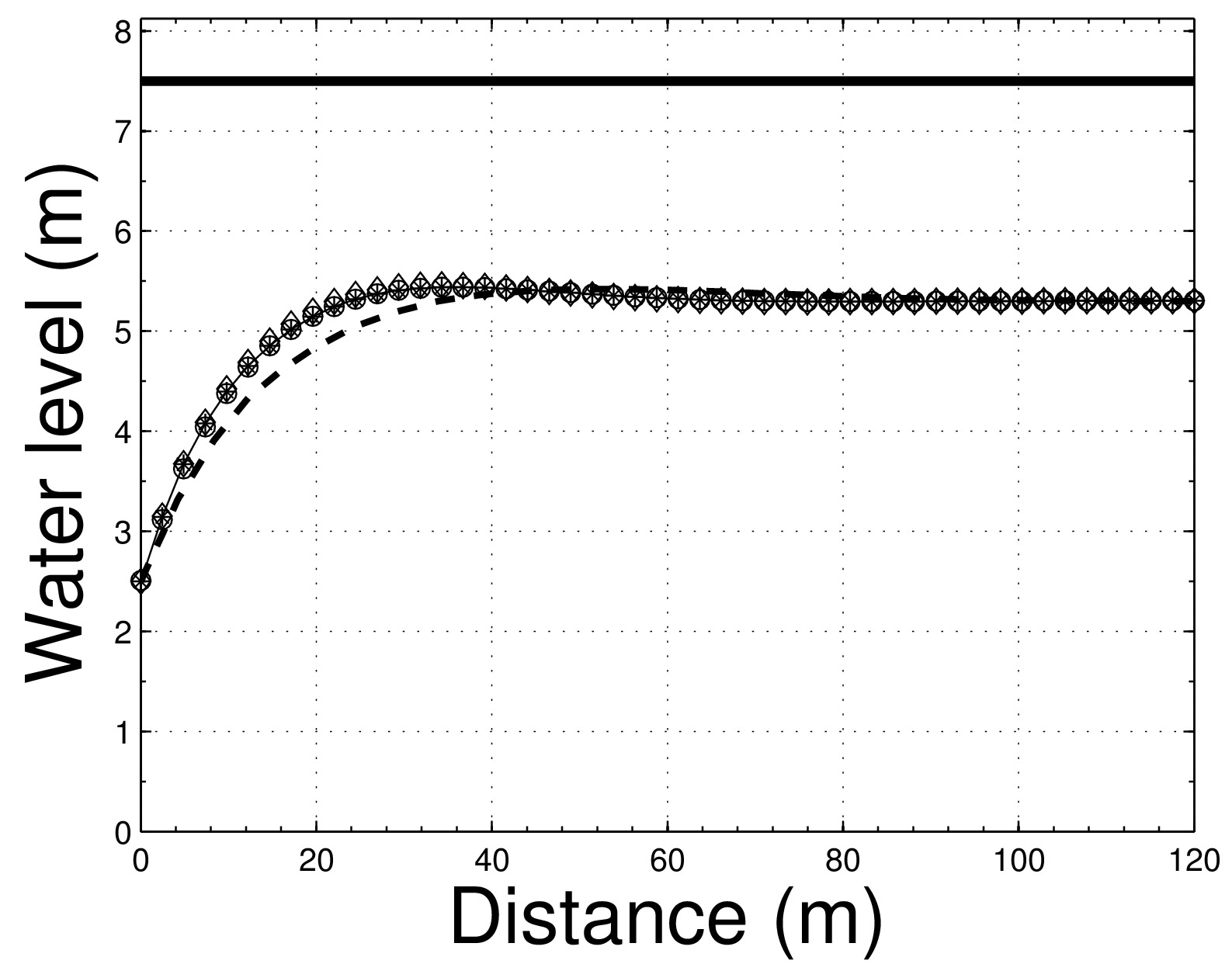}
            \caption[]%
            {{\small t=6 hrs}}
            \label{fig:t_31}
        \end{subfigure}
                \begin{subfigure}[b]{0.6\textwidth}
            \centering
            \includegraphics[width=0.5\textwidth]{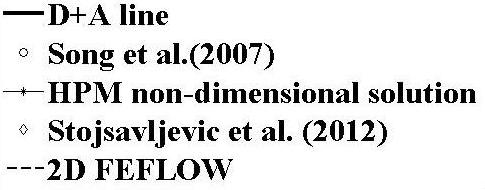}
        \end{subfigure}
        \caption[\small Comparison of Homotopy perturbation solution with \cite{Song2007}, \cite{Stojsavljevic2012}, and 2D FEFLOW solution for a tidal wave with timeperiod= 12 hrs, D=5 m, A=2.5 m, $\eta_e=0.3$, $K=0.00089\:m/s$, and $\omega=(2\pi/12)\:/hr$]
        {\small Comparison of Homotopy perturbation solution with \cite{Song2007}, \cite{Stojsavljevic2012}, and 2D FEFLOW solution for a tidal wave with timeperiod= 12 hrs, D=5 m, A=2.5 m, $\eta_e=0.3$, $K=0.00089\:m/s$, and $\omega=(2\pi/12)\:/hr$}
        \label{fig:HPM_Kong1}
\end{figure*}

To investigate the accuracy of the homotopy perturbation solution, the present solution is compared with previous analytical solutions by \cite{Song2007} and \cite{Stojsavljevic2012} along with numerical solution obtained from FEFLOW numerical modelling (Figure \ref{fig:HPM_Kong1}). Analytical solutions by \cite{Song2007} and \cite{Stojsavljevic2012} based on pre-defined perturbation parameter(s).

\begin{figure*}
        \centering
        \begin{subfigure}[b]{0.475\textwidth}
            \centering
            \includegraphics[width=\textwidth]{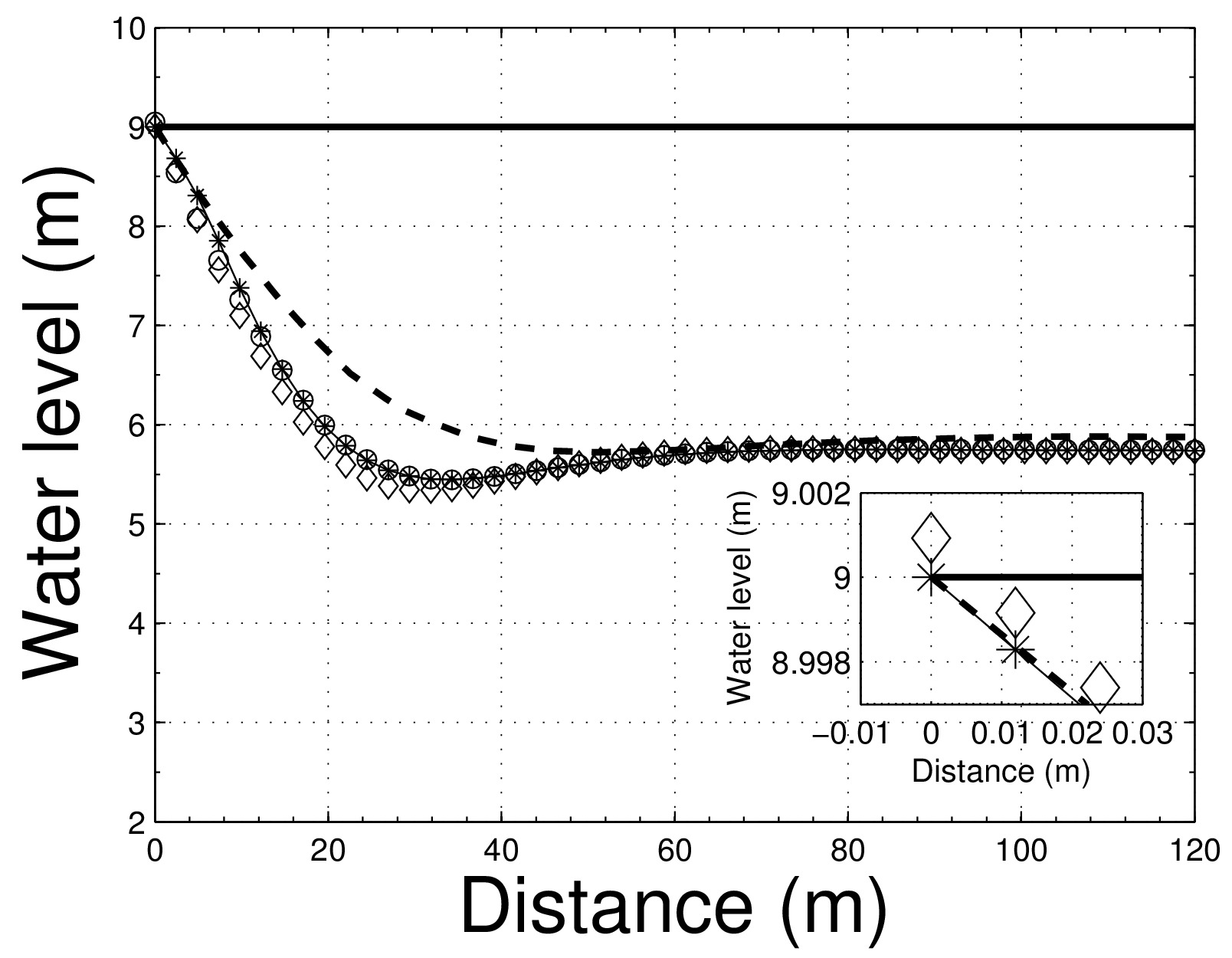}
            \caption[]%
            {{\small t=0 s }}
            \label{fig:t_0}
        \end{subfigure}
        \hfill
        \begin{subfigure}[b]{0.475\textwidth}
            \centering
            \includegraphics[width=\textwidth]{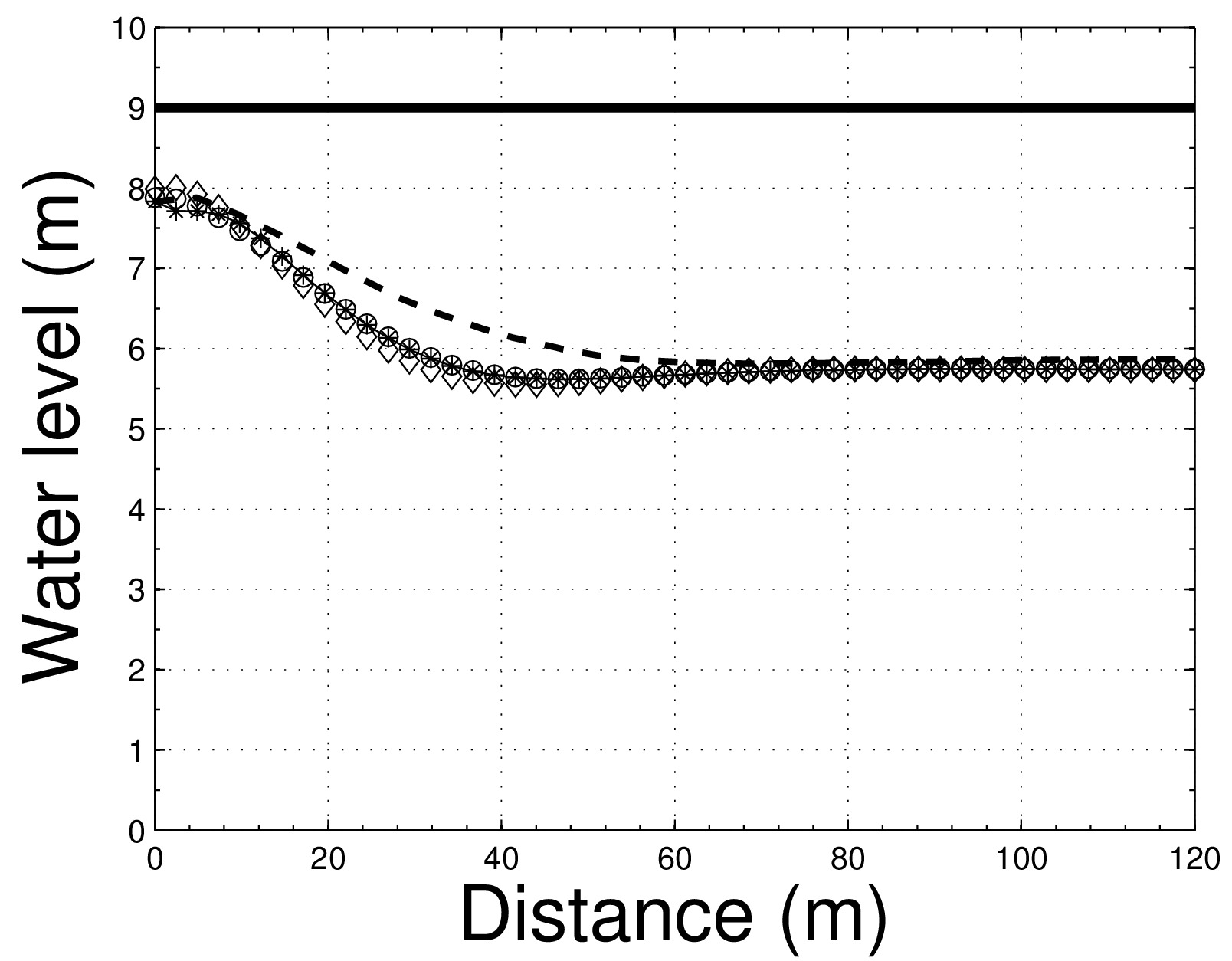}
            \caption[]%
            {{\small t=1.5 hrs  }}
            \label{fig:t_8}
        \end{subfigure}
        \vskip\baselineskip
        \begin{subfigure}[b]{0.475\textwidth}
            \centering
            \includegraphics[width=\textwidth]{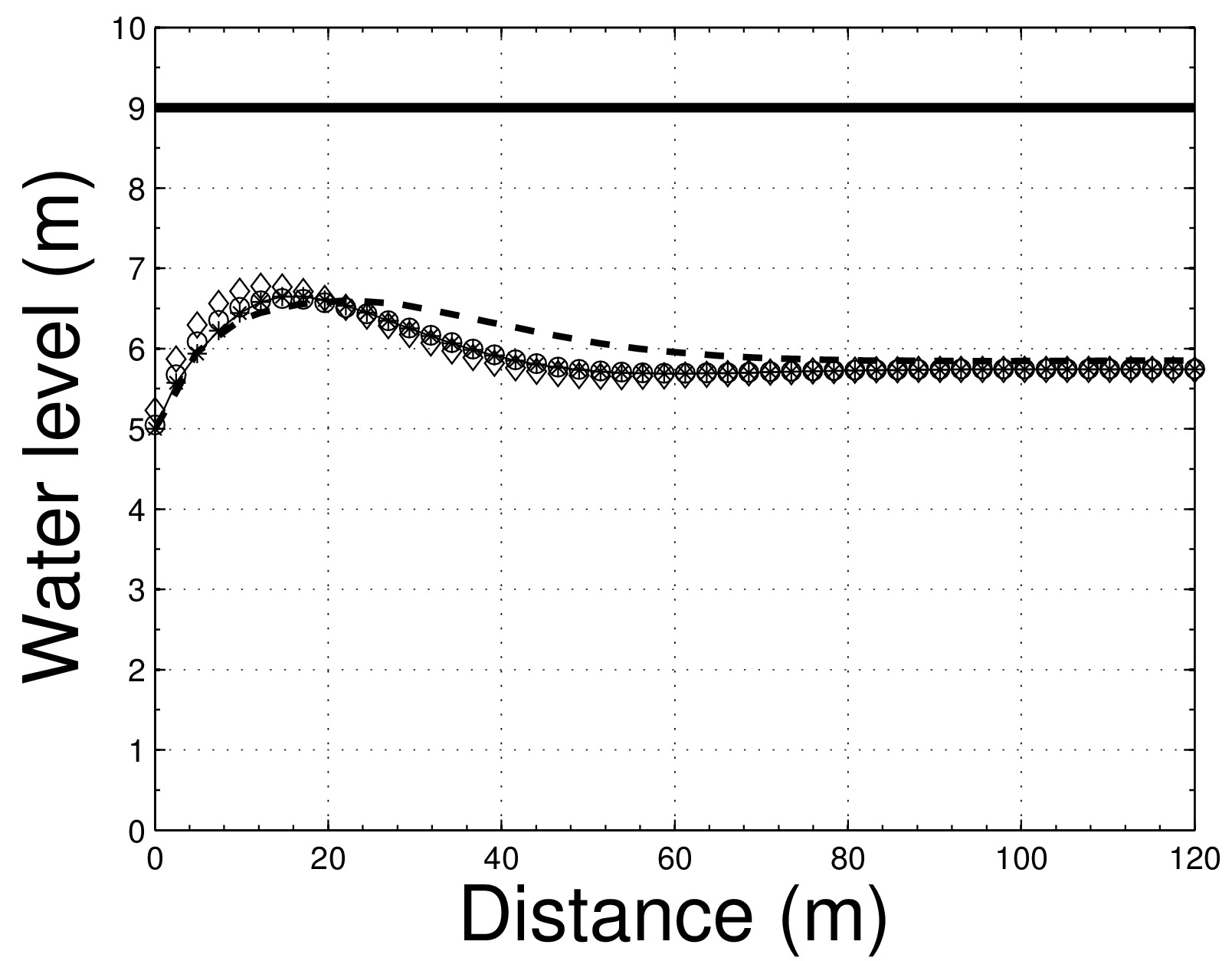}
            \caption[]%
            {{\small t=3 hrs}}
            \label{fig:t_16}
        \end{subfigure}
        \quad
        \begin{subfigure}[b]{0.475\textwidth}
            \centering
            \includegraphics[width=\textwidth]{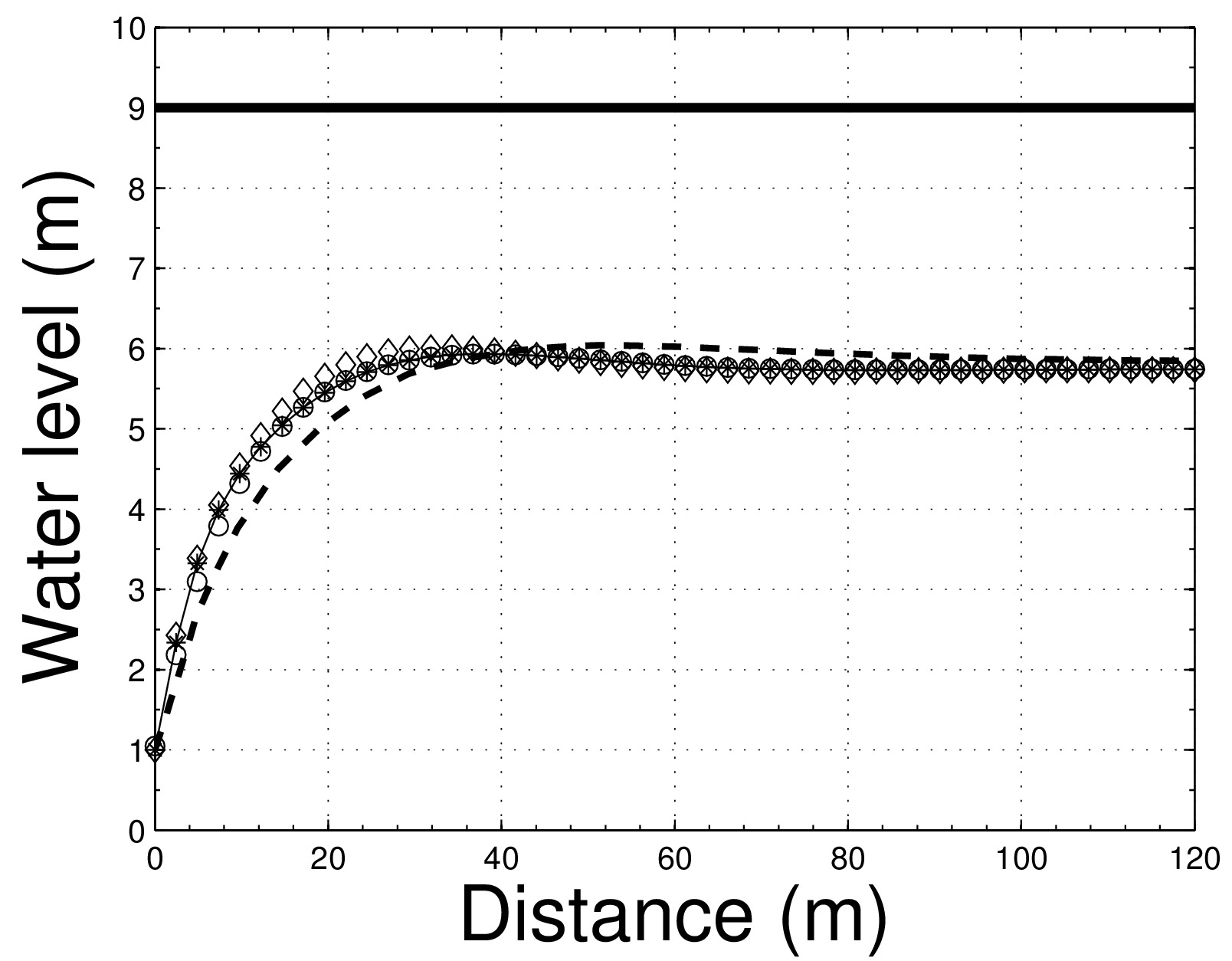}
            \caption[]%
            {{\small t=6 hrs}}
            \label{fig:t_31}
        \end{subfigure}
                \begin{subfigure}[b]{0.6\textwidth}
            \centering
            \includegraphics[width=0.5\textwidth]{legend.jpg}
        \end{subfigure}
        \caption[\small Comparison of Homotopy perturbation solution with \cite{Song2007}, \cite{Stojsavljevic2012}, and 2D FEFLOW solution for a tidal wave with timeperiod= 12 hrs, D=5 m, A=4 m, $\eta_e=0.3$, $K=0.00089\:m/s$, and $\omega=(2\pi/12)\:/hr$ ]
        {\small Comparison of Homotopy perturbation solution with \cite{Song2007}, \cite{Stojsavljevic2012}, and 2D FEFLOW solution for a tidal wave with timeperiod= 12 hrs, D=5 m, A=4 m, $\eta_e=0.3$, $K=0.00089\:m/s$, and $\omega=(2\pi/12)\:/hr$ }
        \label{fig:HPM_Kong2}
\end{figure*}

\cite{Stojsavljevic2012} solution is based on semi-analytical approach by assuming the solution form and ansatz algorithm used to compute the coefficients. The present solution third-order solution for vertical beach face is compared with the \cite{Song2007} second-order solution, and \cite{Stojsavljevic2012} fourth-order solution. The FEFLOW solution is obtained for two-dimensional vertical cross-section with Richards equation. A model domain of $2400\:m \times\: 10 m$ is considered. The number of elements and nodes are $21210$ and $14940$ respectively. The plots are provided for amplitude parameter ($A/D$) values of $0.5$ and $1.0$. The parameters used in the plots are for field scale coastal aquifer. The parameters used for Figure~\ref{fig:HPM_Kong1} are: $A/D=0.5$, $\eta_e=0.3$, $K=0.00089\:m/s$, $D=5\:m$, $A=2.5\:m$, and $\omega=(2\pi/12)\:/hr$. Figure \ref{fig:HPM_Kong2} plotted for $A/D=0.8$ ($D=5\:m$, $A=4\:m$) and other parameters are same as Figure \ref{fig:HPM_Kong1}. The values of the perturbation parameters used in \cite{Stojsavljevic2012} are the following: $\alpha=0.5$ and $\epsilon=0.35$.  The present solution compares well with the previous analytical solutions and also with standard 2D FEFLOW solution \citep{Diersch2013}. The plots are provided with the $A+D$-line to verify the conformity of the solutions at the boundary condition. From the observation of the magnified plot in the Figure \ref{fig:t_0}, \cite{Song2007} and \cite{Stojsavljevic2012} solutions deviate at the seaside boundary condition (Song et al. solution is outside the domain of magnified portion). To stop the generation of secular term in the solution, corrections are introduced by \cite{Song2007} in the perturbation boundary conditions. The non-dimenstional solution is valid for all ranges of $A/D$ values. Moreover, the non-dimensional solution conforms to the seaside boundary condition (Figure \ref{fig:t_0}). Both dimensional and non-dimensional third-order solutions show the presence of the fourth sub-harmonics ($4\omega$) and an additional time invariant term.

\section{Conclusions}

\noindent In the present work, a new analytical solution based on the homotopy perturbation method for tide-induced groundwater level fluctuations is derived. A non-dimensional solution for Boussinesq's equation provided for periodic boundary condition to show the applicability of the homotopy perturbation method. Based on the study, following conclusions can be given:

\begin{itemize}
\item  Analytical solution does not require pre-specified perturbation parameters. The physical parameters involved in the governing equation generates implicit parameter group(s).

\item The secular terms are removed from the solution by using perturbation expansions for the boundary condition.

\item Non-dimensional solution is valid for all ranges $A/D$ values. The solution at the seaside boundary for non-dimensional solutions is consistent with the harmonic cosine function unlike the previous analytical solutions by \cite{Song2007} and \cite{Stojsavljevic2012}.

The analytical solutions can be used for the inverse modelling problem to fit the parameters for wave propagation in coastal aquifers \cite{Raubenheimer1999,Fakir2003}. Analytical solution can also be used to determine coastal aquifer parameters (hydraulic diffusivity, beach slope, and aquifer thickness) and tidal characteristics (amplitude, frequency and phase lag of the tide) with out conducting an aquifer test \cite{Chen2011}

\end{itemize}

\section*{Acknowledgement}

\noindent This study is supported by Science and Engineering Research Board (Grant Number: SB/FTP/ETA-0356/2013) under the Department of Science and Technology, Government of India.

\section*{Appendix}

\subsection*{Secular Terms}

\noindent In the perturbation solution process, secular terms being generated due to nonlinearity in the partial differential equations. Secular terms are the terms with unbounded solution for the particular integral. The inhomogeneous partial differential equation can be written as,
\begin{align}
f(\Delta,\Delta') H(X,T)=g(X,T)
\end{align}
\noindent Where, $f(\Delta,\Delta')$ is the function with partial differentials with respect to independent variables (X and T), $\Delta$ is the partial differential with respect to $X$ (=$\partial/\partial X$), $\Delta'$ is the partial differential with respect to $T$ (=$\partial/\partial T$), $H(X,T)$ is the dependent variable and $g(X,T)$ is inhomogeneous term.

\subsection*{Inhomogeneous Terms}
 \label{Iterms}
\noindent The inhomogeneous terms for the third-order non-dimensional perturbation equations can be written as,
\begin{align}
&J_3\nonumber\\
&=\frac{A^4}{D_{\infty}^4}\left\{\frac{1}{4}\:e^{-2 X}+\frac{59}{20}\:e^{-2\sqrt{2}X}-\frac{3}{10}\:e^{-\sqrt{2}X}\right.\nonumber\\
&\quad-\frac{1}{4}(1+\sqrt{2})\:e^{-(1+\sqrt{2})X}\:\cos[2 T-X]\nonumber\\
&\quad+\frac{\sqrt{3}}{16}(-2+3\sqrt{2})\:e^{-\frac{(1+\sqrt{3})}{\sqrt{2}}X}\:\cos[2 T-\frac{(-1+\sqrt{3}}{\sqrt{2}}X]\nonumber\\
&\quad+\frac{25}{8}\:e^{-2\sqrt{2} X}\:\cos[2 T-\sqrt{2 } X]\nonumber\\
&\quad-\frac{1}{40}(-8+5\sqrt{2})\:e^{-\sqrt{2} X}\:\cos[2 T-\sqrt{2} X]\nonumber\\
&\quad-\frac{17}{8\sqrt{2}}\:e^{-(1+\sqrt{2}) X}\:\cos[(1-\sqrt{2}) X]\nonumber\\
&\quad-\frac{1}{8}(6+5\sqrt{2})\:e^{-(1+\sqrt{2}) X}\:\cos[2 T-X]\nonumber\\
&\quad-\frac{1}{2}\:e^{-2 X}\:\sin[4 T-2 X]\nonumber\\
&\quad+\frac{1}{4}(5+4\sqrt{2})\:e^{-(1+\sqrt{2})X}\:\sin[2 T-X]\nonumber\\
&\quad-\frac{1}{16}(-2+3\sqrt{2})(2+\sqrt{3})\:e^{-\frac{(1+\sqrt{3})}{\sqrt{2}}X}\:\sin[4 T-\frac{(1+\sqrt{3})}{\sqrt{2}}X]\nonumber\\
&\quad-\frac{1}{16}(-2+3\sqrt{2})\:e^{-\frac{(1+\sqrt{3})}{\sqrt{2}}X}\:\sin[2 T-\frac{(-1+\sqrt{3})}{\sqrt{2}}X]\nonumber\\
&\quad-4\:e^{-2\sqrt{2} X}\sin[4 T-2\sqrt{2} X]\nonumber\\
&\quad+\frac{1}{5}\:e^{-2\sqrt{2}X}\:(-25+3\:e^{\sqrt{2}X})\:\sin[2 T-\sqrt{2} X]\nonumber\\
&\quad+\frac{1}{16}(36+25\sqrt{2})\:e^{-(1+\sqrt{2})X}\:\sin[4 T-(1+\sqrt{2})X]\nonumber\\
&\quad+\frac{1}{4}(4+3\sqrt{2})\:e^{-(1+\sqrt{2})X}\:\sin[2 T-X]\nonumber\\
&\quad\left.+\frac{1}{4}\:e^{-(1+\sqrt{2})X}\:\sin[(1-\sqrt{2})X]\right\}\label{eqn:J3}
\end{align}
\noindent Solution for third-order equation ($H_3$) is given by,
\begin{align}
&H_{3}(X,T)\nonumber\\
&=\frac{A^4}{D_{\infty}^4}\left\{\frac{1}{160}(-10\:e^{-2 X}-59\:e^{-2\sqrt{2}X}+24\:e^{-\sqrt{2}X}) \right.\nonumber\\
&\quad-\frac{1}{8}\:e^{-2 X}\:\cos[4 T-2 X]\nonumber\\
&\quad-\frac{(205+223\sqrt{2})}{480(5+2\sqrt{2})}\:e^{-X}\:\cos[2 T-X]\nonumber\\
&\quad+\frac{(38+25\sqrt{2})}{80+32\sqrt{2}}\:e^{-(1+\sqrt{2})X}\:\cos[2 T-X]\nonumber\\
&\quad-\frac{1}{3}\:e^{-2\sqrt{2}X}\:\cos[4 T-2\sqrt{2} X]-\frac{65}{96}\:e^{-2\sqrt{2}X}\:\cos[2 T-2\sqrt{2} X]\nonumber\\
&\quad+\frac{3}{10}\:e^{-\sqrt{2}X}\:\cos[2 T-\sqrt{2}X]\nonumber\\
&\quad+\frac{(29-121\sqrt{2}+14\sqrt{3}+77\sqrt{6})}{336(5+2\sqrt{2})}\:e^{-\sqrt{2}X}\:\cos[4 T-\sqrt{2}X]\nonumber\\
&\quad+\frac{1}{4}\:e^{-(1+\sqrt{2})X}\:\cos[(1-\sqrt{2})X]\nonumber\\
&\quad+\frac{(508+363\sqrt{2})}{224(5+2\sqrt{2})}\:e^{-(1+\sqrt{2})X}\:\cos[4 T-(1+\sqrt{2})X]\nonumber\\
&\quad-\frac{(2+11\sqrt{2})}{32(5+2\sqrt{2})}\:e^{-\frac{(1+\sqrt{3})}{\sqrt{2}}X}\:\cos[2 T-\frac{(-1+\sqrt{3})}{\sqrt{2}}X]\nonumber\\
&\quad-\frac{(2+11\sqrt{2})(3+2\sqrt{3})}{96(5+2\sqrt{2})}\:e^{-\frac{(1+\sqrt{3})}{\sqrt{2}}X}\:\cos[4 T-\frac{(1+\sqrt{3})}{\sqrt{2}}X]\nonumber\\
&\quad-\frac{(-105-96\sqrt{2}+10\sqrt{3}+55\sqrt{6})}{480(5+2\sqrt{2})}\:e^{-X}\:\sin[2 T-X]\nonumber\\
&\quad-\frac{(6+5\sqrt{2})}{8(5+2\sqrt{2})}\:e^{-(1+\sqrt{2})X}\:\sin[2 T-X]+\frac{5}{32}\:e^{-2\sqrt{2}X}\:\sin[2 T-\sqrt{2}X]\nonumber\\
&\quad+\frac{(-20+9\sqrt{2})}{80(5+2\sqrt{2})}\:e^{-\sqrt{2}X}\:\sin[2 T-\sqrt{2}X]\nonumber\\
&\quad+\frac{(4+5\sqrt{2})}{32(5+2\sqrt{2})}\:e^{-(1+\sqrt{2})X}\:\sin[(1-\sqrt{2})X]\nonumber\\
&\quad\left.+\frac{(2+11\sqrt{2})}{32\sqrt{3}(5+2\sqrt{2})}\:e^{-\frac{(1+\sqrt{3})}{\sqrt{2}}X}\:\sin[2 T-\frac{(-1+\sqrt{3})}{\sqrt{2}}X]\right\}
\end{align}
\noindent By applying left boundary condition for the requirement of no secular term in fourth-order inhomogeneous term $b_3$ can be given as,
\begin{align}
b_3=-\frac{A^4}{32 D_{\infty}^4}
\end{align}

\newpage
%\listoffigures
%\listoftables
\newpage
\begin{figure}
\centering
    \includegraphics[width=\linewidth]{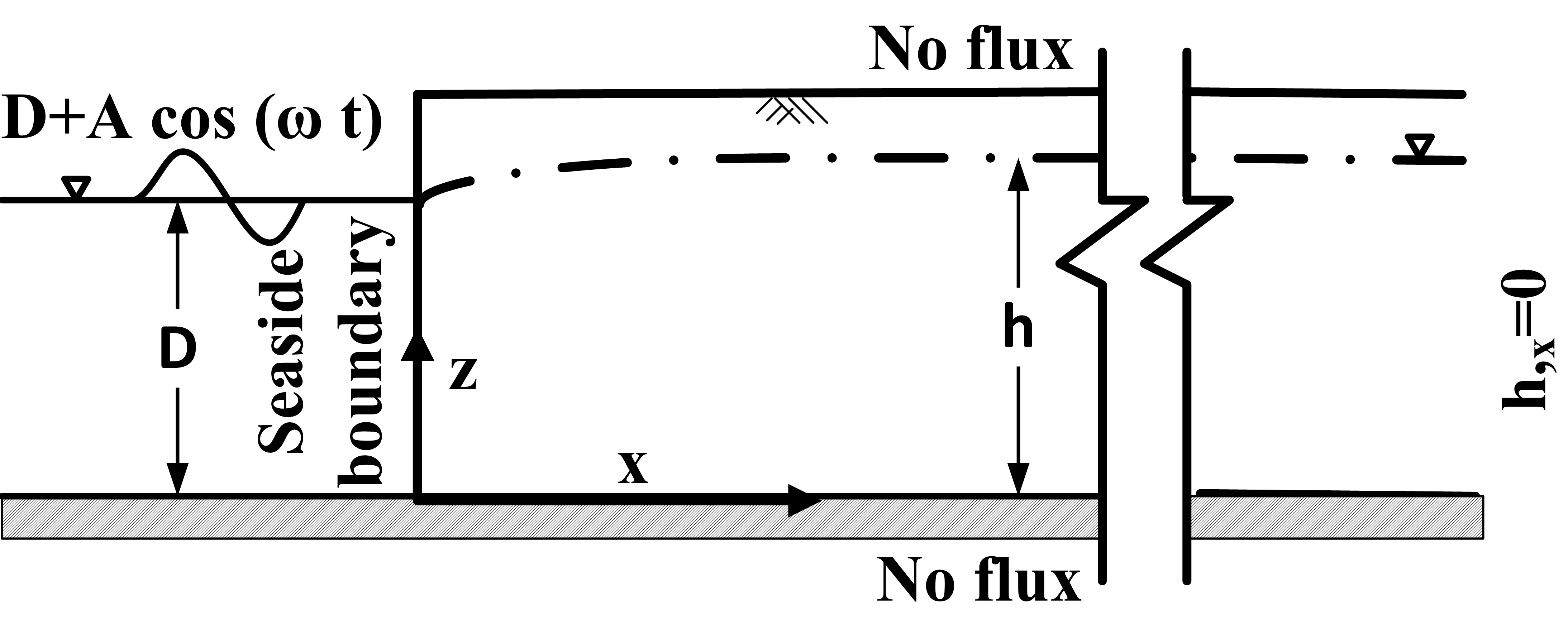}
    \caption{ Definition sketch for tidal influence on groundwater table }
    \label{fig:HPM_sketch}
\end{figure}

\newpage

\newpage

\end{document}